\documentclass[11pt]{article}
\usepackage{color}
\usepackage{multicol}
\definecolor{rossos}{cmyk}{0,1,1,0.55}

\definecolor{rossoc}{cmyk}{0,1,1,0.2}
\definecolor{blu}{cmyk}{1,1,0,0.3}
\definecolor{blus}{cmyk}{1,1,0,0.6}

\font\tenrsfs=rsfs10
\font\sevenrsfs=rsfs7
\font\fiversfs=rsfs5
\newfam\rsfsfam
\textfont\rsfsfam=\tenrsfs
\scriptfont\rsfsfam=\sevenrsfs
\scriptscriptfont\rsfsfam=\fiversfs
\def\mathscr#1{{\fam\rsfsfam\relax#1}}

\def\Amp{\mathscr{M}}
\newcommand{\MeV}{\,\hbox{MeV}}
\newcommand{\GeV}{\,\hbox{GeV}}
\newcommand{\fig}[1]{~{\rm \ref{fig:#1}}}

\def\circa#1{\,\raise.3ex\hbox{$#1$\kern-.75em\lower1ex\hbox{$\sim$}}\,}
\def\Amp{{\cal M}}

\usepackage{graphicx}
\begin{document}
\centerline{astro-ph/0302055\hfill
 LNGS/TH--02/03\hfill
IFUP--TH/2003-2 }

\vspace{1cm}

\centerline{\LARGE\bf\color{rossos} Precise quasielastic neutrino/nucleon cross section}
\vspace{1cm}
\centerline{\bf Alessandro Strumia$^a$ \rm and \bf Francesco Vissani$^b$}
\vspace{0.5cm}

\centerline{$^a$ \em Dipartimento di Fisica dell'Universit\`a di 
Pisa e INFN, Italia}
\centerline{$^b$ \em INFN, Laboratori Nazionali del Gran Sasso, 
Theory Group, Italia}
\vspace{1cm}
{\color{blus}
\centerline{\large\bf Abstract}\begin{quote}\large\indent Quasielastic 
antineutrino/proton and neutrino/neutron
scatterings can be well approximated by simple formul\ae,
valid around MeV or GeV energies.
We obtain a single expression valid in the whole range, 
and discuss its relevance for studies of supernova neutrinos,
which reach intermediate energies.

\end{quote}}

\section{Introduction}
The quasielastic reaction $\bar\nu_e p\to e^+ n$ 
(inverse beta decay) is of special relevance 
for studies of $\bar\nu_e$ at sub-GeV energies~\cite{Cei}.
Indeed, (1)~its cross section 
is relatively large\footnote{It is interesting to recall that the first authors who 
computed it~\cite{Bethe} concluded:
`{\em This meant that one obviously would never be able to see a neutrino.'}}, 
$ \sim G_{\rm F}^2 E_\nu^2$;
(2)~it can be accurately computed;
(3)~it  has a low threshold $E_\nu > 1.806\MeV \approx m_e+m_n - m_p$;
(4)~the measurable $e^+$ energy is strongly correlated with the 
$\bar\nu_e$ energy;
(5)~materials rich in free protons are cheap
(e.g.\  water, hydrocarbon) and this permits to build large detectors;
(6)~in scintillators, it is possible to tag both the $e^+$ and the 
neutron, reducing the  backgrounds.
These features are not shared by other reactions, 
e.g.\  (1), (4) and (6) do not hold for $\nu e\to \nu e$  (since $m_e \ll m_p$); 
$\bar\nu_e \,{}^{16}{\rm O} \to e^+ {}^{16}{\rm N}$
has high threshold $E_\nu >11.4\MeV$ and it is theoretically 
much less clean.

The cross sections for quasielastic neutrino 
and antineutrino scattering on neutron or proton have 
been calculated long ago for  
applications at high energy  $E_\nu \sim $ GeV (e.g.\ accelerators).
The reference formul\ae{} for these applications
are eqs.~(3.18) and (3.22)
in the review of Llewellyn-Smith \cite{lls}, 
routinely used, for instance, in analyses of
atmospheric neutrinos.
Vogel and Beacom \cite{vb} remarked 
that effects suppressed by $(m_n-m_p)/E_\nu$
have been neglected in eq.~(3.22), 
which is a bad approximation at low energy $E_\nu \sim$ MeV
(e.g.\ $\bar{\nu}_e$ generated by nuclear reactors or supernov\ae{}).
These authors proposed theirs  eqs.~(12) and (13) which,
neglecting effects suppressed by powers of $E_\nu/m_p$,
 provide  a simple and precise description of neutrino/nucleon 
scattering at low energy. In this paper, we derive a 
unique formula valid at low, intermediate and high energies,
without assuming $m_n - m_p \ll E_\nu$ nor $E_\nu \ll m_p$.
We will compare our cross section with other 
approximations of common use, and discuss 
the range of energies where they are reliable.
Our findings are relevant in particular for 
(1)~detection  of supernova $\bar{\nu}_e$, that 
can have an energy between few MeV up to about  $60$ MeV; 
(2)~transport of $e^\pm$, $\nu_e$ and $\bar{\nu}_e$ 
inside the collapsing star, where the effective 
temperatures are higher and neutrino energies can reach 200 MeV.

The cross-section is computed in section~\ref{fu} and discussed in section~\ref{discuss}.

\section{Cross section} \label{fu}
We write explicit expressions 
for inverse beta decay (IBD), namely 
quasielastic scattering of 
electron antineutrino on proton:
\begin{equation}
\bar{\nu}_e(p_\nu) + p (p_p)\to e^+(p_e) + n (p_n)\mbox{\hskip1cm IBD reaction.}
\end{equation}
We indicate how to obtain the corresponding expressions for
other antineutrinos $\bar\nu_\mu$ and $\bar\nu_\tau$, 
and for quasielastic neutrino/neutron scattering.
Let us define
\begin{equation}
\Delta=m_n-m_p\approx 1.293\mbox{ MeV},\qquad
M=(m_p+m_n)/2\approx 938.9\mbox{ MeV}.
\end{equation}
The 
differential cross section at tree level in the weak interactions, 
averaged (summed) over initial (final) polarizations is:
\begin{equation}
\frac{d \sigma}{d t}= 
\frac{G_{\rm F}^2 \cos\! ^2\theta_{\rm C}}{2 \pi (s-m_p^2)^2}\ 
|\Amp^2|
\end{equation}
where $G_{\rm F} =  1.16637~10^{-5}/\GeV^2$ 
is the Fermi coupling extracted from $\mu$-decay,
$\cos\theta_{\rm C} = 0.9746\pm 0.0008$~\cite{gino} is the cosine of the Cabibbo angle
and $ \Amp$ has the well-known
current-current structure:
\begin{equation}
\Amp=
\bar{v}_{\nu_e}\gamma^a (1-\gamma_5) v_e
\cdot  
\bar{u}_n \left(
f_1 \gamma_a  + g_1 \gamma_a \gamma_5  +
i f_2 \sigma_{ab} \frac{q^b}{2 M}  + g_2 \frac{q_a}{M} \gamma_5 
\right) u_p
\end{equation}
where $q=p_\nu-p_e=p_n-p_p$.
A straightforward calculation yields\footnote{We give some details of the computation.
The polynomial structure in $s-u$
can be easily understood: if we 
express the leptonic tensor $L^{ab}$  in terms of $p_\ell=p_\nu+p_e$ and $q$, 
and the hadronic one $H^{ab}$ in terms of $p_h=p_p+p_n$ and $q$,
it is sufficient  to note that $p_\ell \cdot p_h=s-u$ (while $q^2=t$, 
$q \cdot  p_\ell=-m_e^2$ and  $q \cdot  p_h=m_n^2-m_p^2= 2 M \Delta$). 
The dependence on $\Delta$ of $|\Amp^2|$ 
can be understood by deepening the same argument, namely:
(1)~one decomposes the leptonic tensor $L^{ab}$ in 4 parts:
$p_\ell^a p_\ell^b$, $q^a q^b-q^2 g^{ab}$, $m_e^2 g^{ab}$,
$\epsilon^{abcd} p_{\ell c} q_d$;
(2)~then, one considers the general Lorentz-invariant decomposition
of the hadronic tensor $H(p_h,q)$ (6 parts); 
(3)~finally, one checks explicitly the 24 
contractions, and verifies that $\Delta$ appears 
only in the combinations $\Delta^2$ and $m_e^2 \Delta $.
}:
\begin{equation}
|\Amp^2|=A(t) - (s-u) B(t) + (s-u)^2 C(t)
\label{m2}
\end{equation}
where 
$s=(p_\nu+p_p)^2$, $t=(p_\nu-p_e)^2$, $u=(p_\nu-p_n)^2$
are the usual Mandelstam variables 
and
%  \begin{equation}
%  \begin{array}{l}
%  16\; A= 
%  (t-m_e^2) \begin{array}[t]{l}
%  \left( 4 |f_1^2| (4+t+m_e^2)+4 |g_1^2| (-4+t+m_e^2) \right. \\ \left. 
%  +|f_2^2|(t^2+4 t + 4 m_e^2) + 4 m_e^2 t |g_2^2|  
%  + 8 {\rm Re}[f_1^* f_2] (2 t + m_e^2) + 16 m_e^2 {\rm Re}[g_1^* g_2]\right)
%  \end{array}\\
%  \ \ -  \Delta^2 \!\!  \begin{array}[t]{l}
%  \left(\  (4|f_1^2| + t |f_2^2|))(4+t- m_e^2)+4 |g_1^2|(4-t+ m_e^2) 
%  \right. \\  \left. 
%  +4 m_e^2 |g_2^2| (t-m_e^2) + 8 {\rm Re}[f_1^* f_2] (2 t - m_e^2) 
%  +16 m_e^2 {\rm Re}[g_1^* g_2] \right)
%  -32 m_e^2 \Delta {\rm Re}[g_1^*(f_1+f_2)]
%  \end{array} 
%  \\[4ex]
%  16\; B= 16 t {\rm Re}[g_1^*(f_1+f_2)] +
%  4 m_e^2\Delta\left( |f_2^2|+{\rm Re}[f_1^* f_2+2 g_1^* g_2]\right)\\[1ex]
%  16\; C=4(|f_1^2|+|g_1^2|)- t |f_2^2|.
%  \label{main}
%  \end{array}
%  \end{equation}
\begin{equation}
\begin{array}{l}
16\; A= 
(t-m_e^2) \begin{array}[t]{l}
\bigg[ 4 |f_1^2| (4M^2+t+m_e^2)+4 |g_1^2| (-4M^2+t+m_e^2)  
+|f_2^2|(t^2/M^2+4 t + 4 m_e^2) + \\+4 m_e^2 t |g_2^2|  /M^2
+ 8 {\rm Re}[f_1^* f_2] (2 t + m_e^2) + 16 m_e^2 {\rm Re}[g_1^* g_2]\bigg]
\end{array}\\\ \ -  \Delta^2 \!\!  \begin{array}[t]{l}
\bigg[\  (4|f_1^2| + t |f_2^2|/M^2)(4M^2+t- m_e^2)+4 |g_1^2|(4M^2-t+ m_e^2) 
+4 m_e^2 |g_2^2| (t-m_e^2)/M^2 +\\+ 8 {\rm Re}[f_1^* f_2] (2 t - m_e^2) 
+16 m_e^2 {\rm Re}[g_1^* g_2] \bigg]
-32 m_e^2M \Delta  {\rm Re}[g_1^*(f_1+f_2)]
\end{array} 
\\[4ex]
16\; B= 16 t {\rm Re}[g_1^*(f_1+f_2)] +
4 m_e^2\Delta\left( |f_2^2|+{\rm Re}[f_1^* f_2+2 g_1^* g_2]\right)/M\\[1ex]
16\; C=4(|f_1^2|+|g_1^2|)- t |f_2^2|/M^2.
\label{main}
\end{array}
\end{equation}
% we introduced the notation $m_e^2=m_e^2$ to have a shorter expression,
% and for the same reason, we 
% In order to get a shorter expression in eq.~(\ref{main}) we
% set $M$ to unity (to restore this factor explicitly, simply 
% note that $A,B,C$ have dimensions $M^4$, $M^2$ and $M^0$).
In the limit $\Delta=0$  eq.~(\ref{main}) reduces to eq.~(3.22) 
of Llewellyn-Smith. 

Since strong interactions are invariant under $T$ and $C$,  
the adimensional form factors $f_i,g_i$ are real functions 
of the transferred 
4-momentum $t=q^2 < 0$.
As discussed in section~\ref{gi} they can be approximated with~\cite{lls}
\begin{equation}
\{f_1,f_2\}=\frac{ \{ 1-{(1+\xi) t}/{4 M^2},\xi \} }{
(1-{t}/{4 M^2}) (1-{t}/{M^2_V})^2},\ \ 
g_1=\frac{g_1(0)}{(1-{t}/{M_A^2})^2},\ \ 
g_2=\frac{2 M^2 g_1}{m_\pi^2-{t}}\label{form}
\end{equation}
where $g_1(0)=-1.270\pm 0.003$~\cite{gino}, 
$M_V^2 = 0.71$ GeV$^2$, $M_A^2\approx 1$ GeV$^2$.
Finally $\xi=  \kappa_p - \kappa_n = 3.706$ is the difference between the proton and
neutron anomalous magnetic moments in units of the nuclear magneton
($\kappa_p = 1.792$, $\kappa_n = -1.913$).

\smallskip

The analogous expressions for $\nu_{e}n$ scattering 
are obtained by exchanging 
$p_\nu\leftrightarrow -p_e$, and $m_p\to m_n$ 
in the flux factor. 
Therefore, $|\Amp^2|$ keeps 
the same expression with $s-u\to u-s$; 
in view of its explicit form, this amounts to just 
replace $B\to -B$ in eq.~(\ref{m2}).
The expressions for ${\nu}_\mu$ or ${\nu}_\tau$ scatterings 
are obtained by replacing $m_e$ with $m_\mu$ or $m_\tau$. 

\bigskip

Our final expression is somewhat involved but allows to
numerically compute any desired integrated cross-section.
It is also useful to present some simple analytic approximation
that is valid at low energies (say, at supernova and reactor 
energies or below).
In fact,  the $\nu_e$, $\bar{\nu}_e$ cross sections  simplify, 
when expanding $|\Amp^2|$ in powers of the parameter: 
\begin{equation}
\varepsilon=E_\nu/m_p
\label{8}
\end{equation}
When we prescribe that $E_e,\Delta,m_e\sim \varepsilon$, 
we get 
$$s-m_p^2\sim \varepsilon,\qquad
s-u\sim \varepsilon,\qquad  t\sim \varepsilon^2.
$$ 
The flux factor in the denominator of $d\sigma/dt$ gives a $1/\epsilon^2$,
and the amplitude $|\Amp^2|$ starts at order $ \varepsilon^2$.
The leading order (LO) approximation in $\varepsilon$ is obtained by 
retaining only
\begin{equation}
\begin{array}{l}
A\simeq M^2(f_1^2-g_1^2) (t-m_e^2)  -M^2\Delta^2 (f_1^2+g_1^2)\\ 
B\simeq  0 \\
C\simeq (f_1^2+g_1^2)/4
\end{array}
\label{LO}
\end{equation} 
%and agrees with the expressions given by Vogel and Beacom~\cite{vb}.
Since $B\simeq  0$ at LO $\nu_e n$ and $\bar{\nu}_e p$ scatterings
are described by the same $\Amp^2$.

A simple approximation which is accurate enough to describe the 
detection of supernova neutrinos events via IBD 
is obtained using the expression of $\Amp^2$ 
accurate up to NLO in $\varepsilon$, given by
\begin{equation}
\begin{array}{l}
A\simeq M^2(f_1^2- g_1^2)(t-m_e^2) - M^2 \Delta^2 (f_1^2+ g_1^2)\ 
- 2 m_e^2 M \Delta g_1 (f_1+f_2) \\ 
B \simeq  t\  g_1 (f_1+f_2)  \\
C\simeq  (f_1^2+g_1^2)/4
\end{array}
\label{10}
\end{equation} 
The $t$ dependence of the form factors contributes at next order (NNLO):
at NLO the form factors are approximated with constants.

\begin{figure*}[t]
\centerline{\mbox{\includegraphics[width=\textwidth]{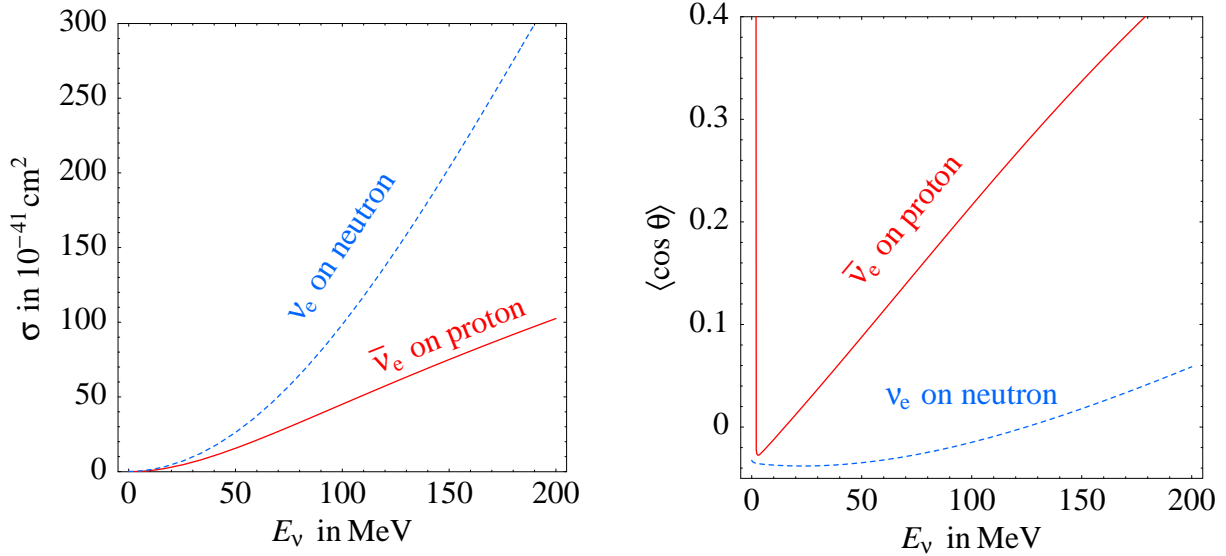}}}
  \caption{\em Fig.~\ref{fig:xsec}a: total cross sections of 
quasielastic scatterings at supernova 
energies.
Fig.~\ref{fig:xsec}b: average cosine of the charged lepton scattering angle
 $\langle \cos\theta\rangle$.
}\label{fig:xsec}
  \end{figure*}
\subsection{Total cross section, $d\sigma/dE_e$ and event numbers \label{22}}
The cross section expressed in 
terms of the neutrino and electron energy in the rest frame of the proton,
$E_\nu$ and $E_e$, is particularly useful.
Inserting
$$s-m_p^2=2 m_p E_\nu,\qquad
s-u=2 m_p (E_\nu+E_e)-m_e^2,\qquad
t = m_n^2-m_p^2-2 m_p (E_\nu -E_e)
%  (m_n - m_p)^2 - 2 m_p (E_\nu - E_e + m_p - m_n)
 $$
it is given by
\begin{equation}
\frac{d\sigma}{d E_e}(E_\nu,E_e)= 2 m_p \frac{d\sigma}{d t}\ \ \ 
\mbox{ if }E_\nu\ge E_{\rm thr}\equiv\frac{(m_n+m_e)^2 -m_p^2}{2 m_p}
\label{thr}
\end{equation}
The allowed values of $E_e$, $E_1 \le E_e \le E_2$, correspond to the possible
scattering angles $\theta^{\rm \scriptscriptstyle CM}$ in the center of mass (CM) frame:
\begin{equation}
E_{1,2}=E_\nu-\delta - \frac{1}{m_p}E_\nu^{\rm \scriptscriptstyle  CM} (E_e^{\rm \scriptscriptstyle  CM}\pm p_e^{\rm \scriptscriptstyle  CM}),\ \ \ 
\mbox{ with } \delta\equiv\frac{m_n^2-m_p^2-m_e^2}{2 m_p}
\label{kin1}
\end{equation}
where the energy and momenta in the CM have the standard expressions:
\begin{equation}
 E_\nu^{\rm \scriptscriptstyle  CM} = {s-m_p^2\over 2\sqrt{s}},\qquad
E_e^{\rm \scriptscriptstyle  CM} = {s-m_n^2+m_e^2\over 2\sqrt{s}},\qquad
p_e^{\rm \scriptscriptstyle  CM} = {\sqrt{[s-(m_n\! -\! m_e)^2] [s-(m_n\! +\! m_e)^2]}\over 2\sqrt{s}}
\end{equation}
For the neutrino reaction, one has just to 
replace $m_n\leftrightarrow m_p$ in all 
formul\ae{} above except in $|\Amp^2|$, where $\Delta$ remains
positive, with the proviso to set $E_{\rm thr}=0$ for $\nu_e$ 
(it would come negative).
The total cross section is plotted in figure \ref{fig:xsec}a and tabulated in table~\ref{tab:tabella}.
We have included also two corrections:
\begin{enumerate}
  \item the radiative corrections to the 
cross section \cite{r1,r2}, well approximated as~\cite{r2}
\begin{equation}
d\sigma(E_\nu,E_e)\to d\sigma(E_\nu,E_e) \ 
\left[ 
1+\frac{\alpha}{\pi} \bigg(6.00 + \frac32\log\frac{m_p}{2 E_e} + 1.2 (\frac{m_e}{E_e})^{1.5}\bigg)
\right]\label{eq:rad}
\end{equation}
where $\alpha$ is the fine-structure constant. 

This one-loop correction 
is valid at $E_\nu \ll m_p$, when 
the final state positron (electron) is the only radiator, 
and in the hypothesis that all the energy $E_e$  in electrons and bremsstrahlung photons is detected.
In general the differential cross section
is given by a more complicated expression which depends on
detection threshold for photons.
The total cross section is still given by eq.~(\ref{eq:rad}), 
if all final state electrons are detected~\cite{r2}.

High-energy electroweak corrections are automatically
included in the values of $g_1(0)$ and $G_{\rm F}$
extracted from low-energy experiments.

% \begin{figure*}[t]
% \centerline{\includegraphics[width=0.7\textwidth]{avec.eps}}
%   \caption{\em Average cosine of the charged lepton scattering angle
% $\langle \cos\theta\rangle$.}\label{fig:ave}
%   \end{figure*}

\begin{table}
  \centering 
$$\begin{array}{|cccc|cccc|cccc|}\hline
E_\nu & \sigma(\bar\nu_e p) & \langle E_e\rangle & \langle \cos\theta\rangle&
E_\nu & \sigma(\bar\nu_e p) & \langle E_e\rangle & \langle \cos\theta\rangle&
E_\nu & \sigma(\bar\nu_e p) & \langle E_e\rangle & \langle \cos\theta\rangle\\ \hline\hline
1.806 & 0  &  -  &  -  &  
8.83 & 0.511  &  7.46  &  -0.015  &  43.2 & 12.1  &  
                  40.2  &  0.070 \\  
2.01 & 0.00351  &  0.719  &  -0.021  &  
9.85 & 0.654  &  8.47  &  -0.013  &  48.2 & 14.7  &  44.8  &  0.083 \\  
2.25 & 0.00735  &  
                                  0.952  &  -0.025  &  11.0 & 0.832  &  9.58  
&  -0.010  &  53.7 & 17.6  &  
                  49.9  &  0.097 \\  
2.51 & 0.0127  &  1.21  &  -0.027  & 
12.3 & 1.05  &  10.8  &  -0.007  & 
 59.9 & 21.0  &  55.6  &  0.113 \\  
2.80 & 0.0202  &  1.50  &  -0.027  &  
13.7 & 1.33  &  12.2  &  -0.003  &  66.9 & 25.0  &  61.8  &  0.131 \\  
3.12 & 0.0304  &  1.82  &  -0.027  &  15.3 & 1.67  &  13.7  &  0.0006  &
74.6 & 29.6  &  68.8  &  0.151 \\  
3.48 & 0.0440  &  2.18  &  -0.027  &  17.0 & 2.09  &  15.5  &  0.005  &  83.2
& 34.8  &  76.5  &  0.173 \\  
3.89 & 0.0619  &  2.58  &  -0.026  &  19.0 & 2.61  &  17.4  &  0.010  &  92.9 
& 40.7  &  85.0  &  0.198 \\  
4.33 & 0.0854  &  3.03  &  -0.025  &  21.2 & 3.24  &  19.5  &  0.015  &  104. & 
47.3  &  94.5  &  0.225 \\  
4.84 & 0.116  &  3.52  &  -0.024  &  23.6 & 4.01  &  21.8  &  0.021  &  116. 
& 54.6  &  105.  &  0.255 \\  
5.40 & 0.155  &  4.08  &  -0.023  &  26.4 &
4.95  &  24.4  &  0.028  &  129. & 62.7  &  117.  &  0.288 \\  
6.02 & 0.205  &  4.69  &  -0.022  &  29.4 & 6.08  &  27.3  &  0.036  &  144. 
& 71.5  &  130.  &  0.323 \\  
6.72 & 0.269  &  5.38  &  -0.020  &  32.8 & 
7.44  &  30.5  &  0.044  &  160. & 81.0  &  144.  &  0.361 \\  
7.49 & 0.349  &  6.15  &  -0.018  &  36.6 & 9.08  &  34.1  &  0.054  &  179. & 91.3  &  161.  &  0.400 \\  
8.36 & 0.451  &  7.00  &  -0.016  &  40.9 & 
11.0  &  38.0  &  0.065  &  200. & 102.  &  179.  &  0.442 \\  
  \hline
\end{array}$$
  \caption{\em Values of $\sigma(\bar\nu_e p\to ne^+)$ in $10^{-41}\,{\rm cm}^2$
and of the average positron energy and cosine of the scattering angle.
All energies are in {\rm MeV}. \label{tab:tabella}}
\end{table}

  \item  We multiply $\sigma(\nu_e n \to p e)$ by
the Sommerfeld  factor that accounts for final-state interactions:
\begin{equation}
F(E_e)=\frac{\eta}{1-\exp(-\eta)}\ ,\ \ 
\mbox{ with }\eta=\frac{2\pi\alpha}{\sqrt{1-m_e^2/E_e^2}}
\end{equation}
\end{enumerate}
Both the radiative corrections and final 
state interactions amount to a $\sim 2$\% correction, 
which can be important for high statistic data samples. 

The formul\ae{} given above can be adapted 
to the case when the nucleon is not at rest (e.g.\ inside a supernova).
Similarly, one can include Fermi blocking-factors 
in final state to account for degenerate electrons sea 
in the supernova.

It is straightforward to compute the average values 
$\langle x\rangle ={\int x\; d\sigma}/{\int d\sigma}$ of interesting  kinematical quantities $x$, like
$
x=E_e,\ E_e^2-\langle E_e\rangle^2,\ \cos\theta$, etc.
We plot in figure 2 and  tabulate in table~\ref{tab:tabella}
the average cosine of the scattering angle of the charged lepton. 
Detecting supernova neutrinos via the IBD reaction,  one could try to
track their source by measuring the direction of positrons, 
that are copiously produced \cite{f,vb} (more discussion below).
Another important quantity is the average lepton 
energy, which can be approximated 
(at better than 1\% below $\sim 100$ MeV) by 
\begin{equation}
\langle E_e\rangle \approx (E_1+E_2)/2=E_\nu-\delta -E_\nu^{\rm \scriptscriptstyle  CM} E_e^{\rm \scriptscriptstyle  CM}/m_p
\end{equation} 
This is a much better approximation than the usual formula
$E_e=E_\nu-\Delta$; it permits to relate $E_e$ with $E_\nu$
incorporating a large part of the effect due to 
the recoil of the nucleon.

It is straightforward to include cuts or efficiencies:
e.g., if the positrons are detected only in the 
window $E_l\le E_e\le E_h$ 
we define 
\begin{equation}
\sigma(E_\nu)=
\int^{E_{2}}_{E_{1}} \! d E_e\; 
\frac{d\sigma}{d E_e}(E_\nu,E_e)\; 
\vartheta(E_e-E_{l}) \vartheta(E_{h}-E_e)
\label{tagliata}
\end{equation}
where $\vartheta(x)$ is the step function, equal to 0 (1) for $x<0$ ($x>0$).
The 
total number of events in the detector and
the distribution in positron energy are given by
\begin{equation}
N_{e^+}= N_p\int d E_\nu 
\frac{d\Phi}{d E_\nu}
\sigma(E_\nu)\qquad
\frac{d N_{e^+}}{dE_e}=  N_p \int  dE_\nu \frac{d\Phi}{d E_\nu} \frac{d\sigma}{dE_e}
\label{numev}
\end{equation}
where $N_p$ is the number of target protons and
$\Phi$ is the $\bar\nu_e$ flux.
We do not have a simple formula for the 
exact limits of integration in $dE_\nu$.
A simple recipe 
is to integrate numerically 
in the larger range
\begin{equation}
E_e+\delta \le E_\nu \le \frac{E_e+\delta}{1-2 (E_e+\delta)/m_p}
\label{recipe}
\end{equation}
obtained replacing $p_e^{\rm \scriptscriptstyle  CM}$ and $E_e^{\rm \scriptscriptstyle  CM}$ with $E_\nu^{\rm \scriptscriptstyle  CM}$ 
in  eq.~(\ref{kin1}), enforcing the correct  
kinematical range by means of $\vartheta$-functions;
the upper limit has to be replaced with 
infinity when the denominator becomes negative.

\subsection{Angular distribution \label{23}}
The cross section differential in the angle $\theta$ between 
the neutrino and the positron  
is obtained from $d\sigma/dt$
after including the Jacobian coming
from $t=m_e^2-2 E_\nu (E_e -p_e \cos\theta)$:
\begin{equation}
\frac{d\sigma}{d\cos\theta} (E_\nu,\cos\theta)
=\frac{p_e \varepsilon}{1+\varepsilon(1  -\frac{E_e}{p_e}\cos\theta)} \ 
\frac{d\sigma}{d E_e},\label{difftheta}
\end{equation}
where $\varepsilon = E_\nu/m_p$.
In eq.~(\ref{difftheta}), $E_e$ and $p_e$ are functions of $E_\nu$ and $\theta$:
\begin{equation}
E_e=\frac{(E_\nu-\delta)(1+\varepsilon) 
+ \varepsilon\cos\theta
\sqrt{(E_\nu-\delta)^2-m_e^2 \kappa}}{\kappa}, \qquad
p_e=\sqrt{E_e^2-m_e^2}
\end{equation}
with 
$\kappa=(1+\varepsilon)^2-(\varepsilon \cos\theta)^2$.
A lepton of low energy cannot be emitted in all directions $\theta=0\div \pi$,
since it is ``dragged'' by the motion of the CM.\footnote{The 
region of ``CM-dragging'' is defined by the condition that the 
CM velocity $\beta$ is larger than $\beta_e^{\rm \scriptscriptstyle  CM}$. 
In this region, the function 
$\theta^{\rm \scriptscriptstyle  CM}(\theta)$ --- or $\theta^{\rm \scriptscriptstyle  CM}(t)$
or $\theta^{\rm \scriptscriptstyle  CM}(E_e)$ --- is not single-valued:
there are two values of $\theta^{\rm \scriptscriptstyle  CM}$ that 
lead to the same  $\theta$. 
This is evident from the relation
$\tan\theta={\sin\theta^{\rm \scriptscriptstyle  CM}}/[\gamma ( \cos\theta^{\rm \scriptscriptstyle  CM} 
+ \beta/\beta_e^{\rm \scriptscriptstyle  CM})]$
where $\gamma=1/\sqrt{1-\beta^2}$.
The region of 'dragging' is visible in the leftmost part 
of figure 2: it is where the curve  ``$\bar{\nu}_e$ on proton'' 
rises suddenly. Exactly at the threshold, 
the positron is emitted only at $\theta=0$,
so that $\langle \cos\theta\rangle \to 1$.} 
The $\tau$, being heavier than the proton, is always emitted in the forward cone.
For $\bar\nu_e$ or $\bar\nu_\mu$ a non trivial allowed range of $\theta$ is obtained
only in a very small region energy range just above the threshold $E_{\rm thr}$
(given in eq.~(\ref{thr})): 
\begin{equation}
E_{\rm thr} \le
E_\nu \le E'_{\rm thr}=E_{\rm thr} + 
m_e \frac{[m_n-(m_p-m_e)]^2}{2 m_p (m_p-m_e)} 
\label{dragthr}
\end{equation}
For the  muon (electron), 
$E'_{\rm thr} - E_{\rm thr} = 0.8 \,{\rm MeV}$ (1 eV).
Above $E'_{\rm thr}$ all scattering angles are allowed. 
Therefore, we can conveniently simplify 
$\bar\nu_e$,  $\bar\nu_\mu$,  scatterings
by replacing $E_{\rm thr}$ with $E_{\rm thr}'$ 
and considering the full range $\theta=0\div \pi$.
This is a very good approximation.

\section{Discussion and applications}\label{discuss}

How important is to use an 
accurate cross section, or in other words,
what is the error that we introduce 
when we use an approximate 
cross section, instead of the correct one? 
The answer, of course, depends on the process
we are considering. 
Here we 
compare our cross section 
with other ones  that can be found in the literature, 
evaluate its uncertainty, and 
finally discuss the specific application 
for supernova $\bar{\nu}_e$ signal.

Here we mention some other possible applications of these results.
A precise cross section can be used in an accurate description of neutrino transport 
in a collapsing supernova (see e.g.~\cite{janka}
for some recent calculations). We 
recall that
neutrino heating (or more generally, the coupling of 
neutrinos with the deepest layers of the mantle)
takes place at energies $E_\nu$ of several tenths of MeV,
and it is regulated by the cross sections that we are 
discussing.
Furthermore, electron neutrinos 
with energies in the range $50-150$ MeV are 
particularly important for the dynamics of neutrino emission:
they are around the Fermi surface of the degenerate 
neutrino distribution, and determine the 
neutrino flux from the inner core.

A precise cross section should 
also be used to interpret searches for $\bar{\nu}_e$ or 
$\bar{\nu}_\mu$  low energy neutrinos from 
the sun (see e.g.~\cite{SKsun}), from gamma-ray-burst or 
other presently unknown cosmic sources
% \footnote{{\bf However, 
% intense sources of low energy neutrinos require very special
% conditions, like those provided by a supernova (a very optically
% thick region around the production site), or rather exotic situations
% like mixing between ordinary and mirror neutrinos.}} 
% NON LO DIREI
\cite{lsd};
or also, in studies of $\nu_\mu$ or $\bar{\nu}_\mu$ 
interactions just above the threshold of 
$\mu$ production (e.g., for large sample of data in 
terrestrial experiments).

The LSND and {\sc Karmen} experiments~\cite{LSNDKarmen} used
a $\bar\nu_\mu$ flux  with a continous spectrum up to
$52.8\MeV$, mainly produced by $\mu^+$ decay at rest,
looking for $\bar\nu_e$ appearance using the $\bar\nu_e p \to e^+ n$ reaction.
The experimental result is controversial.
If such a signal exists, 
a precise computation of the $\bar\nu_e p$ cross-section 
would be necessary to interpret a more precise future experiment.

\subsection{Comparison with various simple approximations}
Table \ref{tab1} shows the percentage difference 
\begin{equation}
\delta \sigma =100\times (1-\sigma'/\sigma)\end{equation}
between our cross section $\sigma$ and 
various approximations $\sigma'$, evaluated at the same input parameters
(percentages below 0.1\% have been set to zero).
The first row shows the neutrino energies at which the 
comparison is made.
In the  upper panel we consider  the $\bar{\nu}_e$ reaction;
in the lower panel the ${\nu}_e$ reaction.

\begin{table}[t]
\centerline{
\begin{tabular}{|ccc|rrrrrrr|}
\hline
\multicolumn{3}{|c|}{$E_\nu$, MeV}  & 2.5 & 5 & 10 & 20 & 40 & 80 & 160 \\   \hline\hline
\multicolumn{10}{|c|}{Percentage difference in $\sigma(\bar\nu_e p\to n \bar{e})$}\\ \hline
1&Na\"{\i}ve    &$\star\star\star$ & $-3.9$ & $-5.8$ & $-9.9$ & $-19$ & $-38$ & $-84$ & $-210$ \\ 
2&Na\"{\i}ve +    &$\star\star\star$ & $0$ & $0.3$ & $-0.2$ & $0.4$ & $0.2$ & $0.5$ & $-0.9$ \\ 
3&Vogel and Beacom   &$\star\star$     &0 & 0 & $0.3$ & 1.2 & 5.6 & 28 & 150 \\
4&NLO in $E_\nu/m_p$  &$\star$ & 0 & 0 & 0 & 0 & 0.1 & 1.5  & 13\\ 
5&Horowitz    &$\star\star$    & $-370$ & $-83$ & $-32$ & $-14$ & $-6.4$ & $-3.0$ & $-1.3$ \\
% 4&Llewellyn-Smith        & $-370$ & $-83$ & $-32$ & $-14$ & $-6.5$ &$-3.0$ &$-1.4$ \\
6&Llewellyn-Smith +  &$\star$     & $-13$ & $-2.1$ & $  -0.5$ & $  -0.1$ &  0 &  0 & 0 \\
7&LS + VB      &$\star$ & $0.5$ & $0.1$ & 0 & 0 & 0 &  0 &0 \\
\hline\hline
\multicolumn{10}{|c|}{Percentage difference in $\sigma(\nu_e n\to pe)$}\\ \hline
1&Na\"{\i}ve   &$\star\star\star$  & $-1.7$ & $-1.5$ & $-1.0$ & $-0.4$ & $0.2$ & $-1.5$ & $-14$ \\
4&NLO in $E_\nu/m_p$  &$\star\star\star$ & 0 & 0 & 0 & $-0.1$ & $-0.5$ & $-2.4$  & $-12$\\
5&Horowitz     &$\star\star$   & $56$ & $37$ & $22$ & $12$ & $6.3$ & $3.2$ & $1.6$ \\
6&Llewellyn-Smith +   &$\star$    & $-3.9$ & $-1.2$ & $-0.3$ & $-0.1$ & 0 & 0 & 0 \\
\hline
\end{tabular}}
\caption{\em 
Percentage difference between our full result and various approximations for $\bar{\nu}_e$
(above) and $\nu_e$ (below) total cross sections.
A negative (positive) sign means that a certain 
cross section is an over(under)-estimate.
It is easy to implement approximations maked with $\star\star\star$, 
while implementing those marked with a $\star$ is not much simpler than performing a full computation.
}
\label{tab1}
\end{table}

\begin{enumerate}
\item The {\bf na\"{\i}ve} low-energy approximation (see e.g.\ \cite{bemporad})
\begin{equation}
\label{naive}
\sigma\approx 9.52\, \times 10^{-44} 
\frac{p_e E_e}{\hbox{MeV}^2}\ \mbox{cm}^2,\qquad
E_e=E_\nu \pm \Delta\hbox{ for }\bar{\nu}_e\hbox{ and }\nu_e,
\end{equation}
obtained by normalizing the leading-order 
result to the neutron lifetime, 
overestimates $\sigma(\bar{\nu}_e p)$ especially at high energy.
It is not recommended for analyses of supernova 
neutrinos, nor for precise studies of reactor $\bar\nu_e$.
The na\"{\i}ve $\sigma(\nu_e n)$, instead, agrees 
well with the exact cross section. 

\item 
A simple approximation which agrees with our full result
within few per-mille for $E_\nu\circa{<}300\MeV$ is
\begin{equation}
\label{naive+}
\sigma(\bar\nu_e p)\approx 10^{-43}\,\mbox{cm}^2~p_e E_e ~E_\nu^{-0.07056+0.02018\ln E_\nu-0.001953\ln^3 E_\nu},
\qquad E_e = E_\nu - \Delta
\end{equation}
where all energies are expressed in MeV.

\item  The low-energy approximation of 
{\bf Vogel and Beacom}~\cite{vb}
(which include first order corrections in $\varepsilon=E_\nu/m_p$, given only for antineutrinos) is very accurate
at low energies ($E_\nu\circa{<}60$ MeV), however underestimates 
the number of supernova IBD neutrino events at highest 
energies by $10$\%.
Higher order terms in $\varepsilon$ happen to be dominant already at $E_\nu \circa{>}135\MeV$,
where the expansion breaks down giving a negative cross-section~\cite{vb,hor}. 

% At higher energies outside its domain of validity
% the LO expansion
% becomes progressively more inadequate and finally produces a negative cross-section~\cite{vb,hor}. 

%  {\em Numbers in our table~\ref{tab1} can be slightly different from~\cite{vb}},
%  where the cross section is expanded in $\epsilon$.
%  For simplicity we instead expand the squared amplitude in $\epsilon$ and
%  treat kinematics exactly,
%  so that some higher order terms are inclded in our LO cross section.

\item The  {\bf NLO} low-energy approximation, defined by eq.~(\ref{10}),
can be used  from low energies up to the energies relevant for
supernova $\bar{\nu}_e$ detection.
(We expand the squared amplitude in $\varepsilon$ but, unlike Vogel and Beacom
we treat kinematics exactly,
so that some higher order terms are included in our NLO cross section).

\item The high-energy approximation of {\bf Horowitz}~\cite{hor},
obtained from the Llewellyn-Smith formul\ae{} setting $m_e = 0$,
 was not tailored 
to be used below $\sim 10$ MeV, and it is not 
precise in the region relevant for supernova neutrino 
detection; however, it is presumably 
adequate to describe supernova neutrino transport.

\item The  {\bf Llewellyn-Smith} high-energy 
approximation, improved adding
$m_n\neq m_p$ in $s,t,u$, but not in $\Amp$ 
is very accurate at all energies relevant 
for supernova neutrinos, failing only at the lowest energies. 
As proved in section~\ref{fu},
this  is a consequence of the absence in $|\Amp^2|$ 
of  corrections of order $\Delta/m_p$.

\item Finally, approximation 6.\ can be improved by including also the dominant
low-energy effects in the amplitude $\Amp$,
as discussed in section~IIB of~\cite{vb}.
With respect to our full result, this amounts to
neglect $\Delta$ and $m_e$ in $A,B,C$, eq.~(\ref{main}),
and reinsert part of them modifying $A\to A - 4M^2\Delta^2 C$.
This approximation agrees with our full result within few per mille.

% Ref.~\cite{vb} suggested match high-energy approximation
% another simple approximation for $\sigma(\bar\nu_e p)$  adequate for supernova $\bar{\nu}_e$ detection: to use the Vogel and Beacom approximation at low energy, 
% and an improved version of Llewellyn-Smith cross section at high energy, 
% as those  proposed in their section II B. 
% Indeed, table~\ref{tab1} shows that an agreement within $1\%$
% with the full result is obtained by 
% changing description at $E_\nu \sim 15 \MeV$.

\end{enumerate}
% Ref.~\cite{vb} suggested another simple approximation for $\sigma(\bar\nu_e p)$  adequate for supernova $\bar{\nu}_e$ detection: to use the Vogel and Beacom approximation at low energy, 
% and an improved version of Llewellyn-Smith cross section at high energy, 
% as those  proposed in their section II B. 
% Indeed, table~\ref{tab1} shows that an agreement within $1\%$
% with the full result is obtained by 
% changing description at $E_\nu \sim 15 \MeV$.

% Furthermore, we have not included
% radiative corrections when comparing
% the cross sections, and used the same form factors.

\subsection{Overall uncertainty}\label{gi}
We now discuss  how accurate our full expressions
for the cross sections are.
All input quantities have been precisely measured, 
except the $t$-dependence of the form factors of eq.~(\ref{form}).
As shown in section~2, it affects the cross sections only at NNLO in $E_\nu/m_p$.
In practice, taking the form factors as constants overestimates the total $\bar\nu_e p$
and $\nu_e n$
cross sections by about $1\%\cdot (E_\nu/40\MeV)^2$.
% {\bf $1.2 y +0.2 y^2 \%$ for antinu,
% $1.1 y^2 \%$ for nu, where $y=E_\nu/40\MeV$}
At low energy this is an excellent approximation, and
the dominant uncertainty is due to $\theta_{\rm C}$ and $g_1(0)$.
The values we adopted 
are extracted from different sources.
The Cabibbo angle $\theta_{\rm C}$ is extracted independently
from $V_{us}$
(measured from $K_{\ell 3}$ decays)
and from $V_{ud}$
(measured from nuclear decays
dominated by vectorial matrix elements,
and from neutron decay).
Since the two determinations agree only within $2.2\sigma$
and theoretical errors play an important r\^ole,
the total error has been conservatively increased~\cite{gino}.

The axial coupling $g_1(0)$ is measured from neutron decay\footnote{Its uncertainty is fully 
correlated with the uncertainty on the $\theta_{\rm C}$ determination from neutron decay, 
since  neutron decay measures the same combination of $g_1(0)$ and $\theta_{\rm C}$ that enters in IBD.}.
Different experimental determinations do not fully agree,
therefore we conservatively increased the error.
Newer measurements, performed with a higher neutron polarization
than older ones, are consistent and agree on
$g_1(0) /f_1(0)= -1.272\pm 0.002$ when older
determinations are discarded --- a value slightly different from the one quoted in section~2.
Isospin-breaking corrections to $f_1(0) =1$ are negligible~\cite{gV}.

In conclusion, at low energy $\sigma(\bar\nu_e p)$ has an overall $0.4\%$
uncertainty, which is adequate for present experiments.
The ratio between the measured and the no-oscillation reactor $\bar{\nu}_e$
flux is $1.01 \pm 2.8\% \hbox{(stat)} \pm 2.7\% \hbox{(syst)}$ at CHOOZ~\cite{CHOOZ}.
The KamLAND collaboration presently has a $6\%$
systematic error on their reactor $\bar{\nu}_e$ data~\cite{k}, 
and aims at reaching the 4\% level in future. 
Concerning supernova $\bar\nu_e$,
experiments should collect few hundreds to several thousand 
of  events, depending on when and where the next 
galactic supernova explosion will occur~\cite{Cei,SNnu,LVD}.

\medskip

We finally discuss the uncertainty introduced by the 
$t$ dependence of the form factors, 
which for the largest supernova $\bar\nu_e$ or $\nu_e$ energies
is not fully negligible.
The SU(2) isospin  symmetry (broken only by small quark masses and by electromagnetism)
 allows  to relate the vector form factors $f_1$ and $f_2$ to measured $eN$ 
cross sections, and conservation of the vector current predicts $f_1(0)=1$.
 The first terms in the expansion in $t$
 $$ f_1(t)  \simeq 1 + 2.5 t/\GeV^2,\qquad
 f_2(t) \simeq \xi [1 + 3.4 t/\GeV^2]$$
 are known with about $1\%$ accuracy~\cite{Mergell}.
The phenomenological parametrization in eq.~(\ref{form}),
partially suggested by theoretical considerations,
reproduces reasonably well the experimental data.
Employing slightly more accurate parametrizations for the vector form factors $f_1$ and $f_2$
(e.g.~\cite{Mergell}) would not improve the accuracy of our results due to the larger uncertainty on
 the axial form factors $g_1$ and $g_2$.

\smallskip

The axial form factor $g_2$ 
(written in eq.~(\ref{form}) in terms of $g_1$, as suggested partial conservation 
of the axial current) 
has a negligible impact on $\nu_e$ and $\bar{\nu}_e$ 
scattering at intermediate energies.
% For heavier leptons, this is not so: for example,
% omitting $g_2$ altogether one overestimates 
% by 15 \% the cross sections $\sigma(\bar{\nu}_\mu p)$ 
% at $E_\nu=150$ MeV.
% New accurate measurements \cite{capt} 
% of muon capture on protons ($t\approx -m_\mu^2$)
% could obtain a precise value of $g_2(0)$: 
% present agreement with theory is poor.
% Note however that the question about 
% $g_2(0)$ is not completely
% disconnected from the $t$-dependence of $g_1$ at low energies;
% for instance, if one sets $t=0$ in the axial form factors,
% one overstimates by 8 \% the cross sections 
% $\sigma(\bar{\nu}_\mu p)$ at 
% 150 MeV. ~\cite{capt} }
Thus, the main uncertainty comes 
from $g_1$.
Vector meson dominance would suggest 
$g_1=g_1(0)/(1-t/M_{A_1}^2)$ where
$M_{A_1} = (1.23\pm 0.04)\GeV$
is the mass of the pseudo-vector
that, mixing with the $W$ boson, 
should give the dominant effect.
The parametrization of $g_1$ adopted in eq.~(\ref{form}) is obtained by
squaring  $(1-t/M_A^2)$ and leaving $M_A$ as a free parameter
because this provides a better fit of data.
Lattice simulations do not contradict 
this hypothesis within the present theoretical errors, that 
are however still large \cite{s}.
Assuming that the parametrization is correct, one obtains
\begin{enumerate}
  \item $M_A = (1.069\pm 0.016)\GeV$~\cite{review} from $\pi$ electroproduction on nucleons ($\gamma^* N\to \pi N'$).
The presence of $\pi$ makes this probe experimentally precise but theoretically uncertain,
as we are interested in $\langle n | \cdots |p\rangle$, not in $\langle \pi n|\cdots | p\rangle$.
  \item $M_A = (1.026 \pm 0.021) \GeV$~\cite{mA,review} from global fits of $\nu N$ scattering data.
Most of the data are not taken at $|t|\ll \GeV^2$, where the parametrization of $g_1$ is certainly justified.
%It would be interestin fit of low-energy $\nu N$ data could

\item $\mu$ capture on $p$ could provide a clean and direct probe of $g_2$ at $t = -0.88 m_\mu^2$.
The present experimental data~\cite{prc} hint to a value about $50\%$ larger than
the one we assume\footnote{Such an increase of  $g_2$ has some impact on $\nu_\mu$ and $\bar\nu_\mu$ scattering,
and would decrease $\sigma(\bar\nu_\mu p)$ by 
$3.8 \% \times (150 \MeV/E_\nu)^3$.}, but their interpretation is under debate~\cite{review}.
New data should clarify the issue~\cite{capt}.

\end{enumerate}
%   A dedicated fit of sub-GeV neutrino scattering data 
%   (or a precise QCD prediction e.g. from lattice simulation~\cite{LATTICE})
%   %or improved determinations from pion electroproduction~\cite{reviewRA})
%   % COME IMPROVI LE INCERTEZZE TEORICHE?
%   could clarify the issue.
The above discussion shows why it is difficult to assess the uncertainty on $g_1$ and $g_2$.
Optimistically assuming that  1.~or 2.~is right, it is negligible.
On the other side, a pessimistic estimate can be obtained by using $M_{A_1}$ in place of $M_A$:
 the total $\bar\nu_e p$ 
cross section increases by $0.4\% (E_\nu/50\MeV)^2$ 
for $E_\nu\circa{<}200\MeV$.
%We take this difference as an estimate of the theoretical error.
The shift remains relatively small because, as shown in section~2, the $t$-dependence of the form factors
affects $\bar\nu_e p$ only at NNLO in $E_\nu/m_p$.

\subsection{Search for relic supernova neutrinos}
Here we discuss the application of our cross 
section to the search for 
neutrino radiation  
from past core-collapse supernov\ae.
We focus on the energy range $E_e>20$ MeV 
which is particularly suited for this type of search, 
since it contains a large part of the expected signal
and it is largely free from background (for 
an experimental proof, see \cite{relic}).
We assume that the energy distribution of supernova $\bar{\nu}_e$ 
in the detectors is described by the following differential flux (see e.g.~\cite{janka,SNnu}):
\begin{equation}
\frac{d\Phi}{d E_\nu} = \frac{E_{\rm tot}}{4\pi D^2} \bigg[
\frac{f_e}{T_e^2}\; \cos^2\theta_{12}\; n\bigg(\frac{E_\nu}{T_e}\bigg) + 
\frac{f_{\mu,\tau}}{T_{\mu,\tau}^2}\; \sin^2\theta_{12}\; n\bigg(\frac{E_\nu}{T_{\mu,\tau}}\bigg)\bigg]
\label{fluxsn}
\end{equation}
where
{\em 1)}~$\theta_{12}\approx 34^\circ$ describes 
`solar' oscillations in supernova mantle;
{\em 2)}~$n(x)= {120}x^2/{7\pi^4}[1+e^x]$ is the
Fermi-Dirac occupation factor;
{\em 3)}~$f_e\approx 1/5$ and $f_{\mu,\tau}\approx 1/7$  are 
the energy shares in $\bar{\nu}_e$ and in $\bar\nu_{\mu,\tau}$; 
{\em 4)}~$T_e\sim\hbox{few MeV}$ is the temperature of $\bar{\nu}_e$ (before oscillations) and 
{\em 5)}~$T_{\mu,\tau}$ is the temperature of $\bar{\nu}_{\mu,\tau}$, that we 
assume to be $T_{\mu,\tau} \approx 1.3\times T_e$;
{\em 6)} $E_{\rm tot}$ is the total energy released in neutrinos, 
$D$ the distance of the supernova.
We neglect various effects that would not change our conclusions, 
like `pinching'  of the spectrum (i.e.\ non-thermal effects), 
eventual oscillations in the Earth, 
existence of other detection reactions
(we focus on  detectors with a large fraction of free protons
like water \v{C}erenkov or scintillators), 
precise definition of the search  window (which depends on the background level), 
detector efficiencies, 
other oscillations eventually induced by a non-vanishing $\theta_{13}$,
etc.

We compare the expected number of events $N_{e^+}$  in the energy range $E_e=[20, 40]$ MeV
using (a) the cross section of eq.~(\ref{tagliata})
and (b) the na\"{\i}ve (and much used) cross 
section in eq.~(\ref{naive}). 
The latter leads to an overestimate:
\begin{equation}
100\times \left(1-
\frac{N_{e^+}({\rm b})}{N_{e^+}({\rm a})}\right)
\approx -25\% + 2.4\% \times x,  
\qquad \mbox{where}\qquad x=\frac{T_e-4.5\MeV}{\MeV}.
\label{acc}
\end{equation} 
The Super-Kamiokande collaboration
presently gives the dominant bound on the relic $\bar\nu_e$ flux,
constrained to be less than
$1.2\bar\nu_e/\hbox{cm}^2\hbox{s}$ at $90\%$ CL
for $E_\nu> 19.3\MeV$~\cite{relic} ---
slightly above the flux expected by cosmological models~\cite{relic2}.
Since this bound has been obtained assuming the na\"{\i}ve $\bar\nu_e p$ cross section,
eq.~(\ref{acc}) suggests that it could be weakened by $\sim 20\%$
(a reanalysis of the data with a precise cross section would allow to be more precise).
  Similar considerations apply to certain calculations of the expected 
  event number from next galactic supernova (e.g.\ ref.~\cite{LVD}) and to
estimates of the total energy $E_{\rm tot}$ from SN1987A neutrino data (e.g.\ ref.~\cite{ll}),
although the low energy cuts are typically smaller than 20 MeV 
and therefore the inaccuracy is less severe.

\subsection{Low energy atmospheric neutrinos as a background}
Low energy atmospheric neutrinos constitute a background
for searches of relic supernova neutrinos.
At Super-Kamiokande \cite{relic}, this background 
is mostly due to muons produced below the 
\v{C}erenkov threshold in water ($E_\mu<160$ MeV),   
whose decay produces a positron.
Assuming that at low energy the $\bar{\nu}_\mu$ flux 
scales as $\sim 1/E_\nu$~\cite{gb},
the cross section with $m_p=m_n=M$ overestimates  
the background only by 1.5\%. 
This is smaller than
what suggested by a simple minded estimate,
%e.g.,\ at $E_\nu=130$ MeV, the overestimate is  5.5\%
performed looking at the total cross section
without taking into account that the condition 
that the muon is produced below the \v{C}erenkov 
threshold {\em does depend} on $\Delta =m_n-m_p$.
Similar considerations
apply to the $\bar{\nu}_e$ atmospheric neutrino 
background: e.g., 
the use of the Llewellyn-Smith
cross section for $E_e=[50,80]$ MeV 
introduces a 2.4\% overestimate.

\begin{figure*}[t]
\centerline{\hspace{-7mm}
\includegraphics[width=0.45\textwidth]{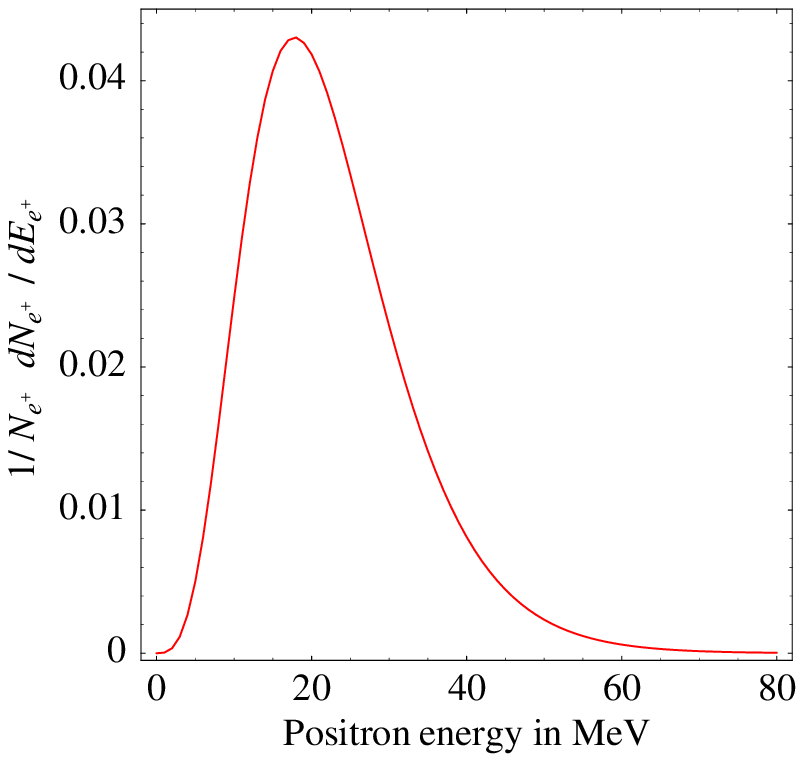}\hspace{1cm}
\includegraphics[width=0.45\textwidth]{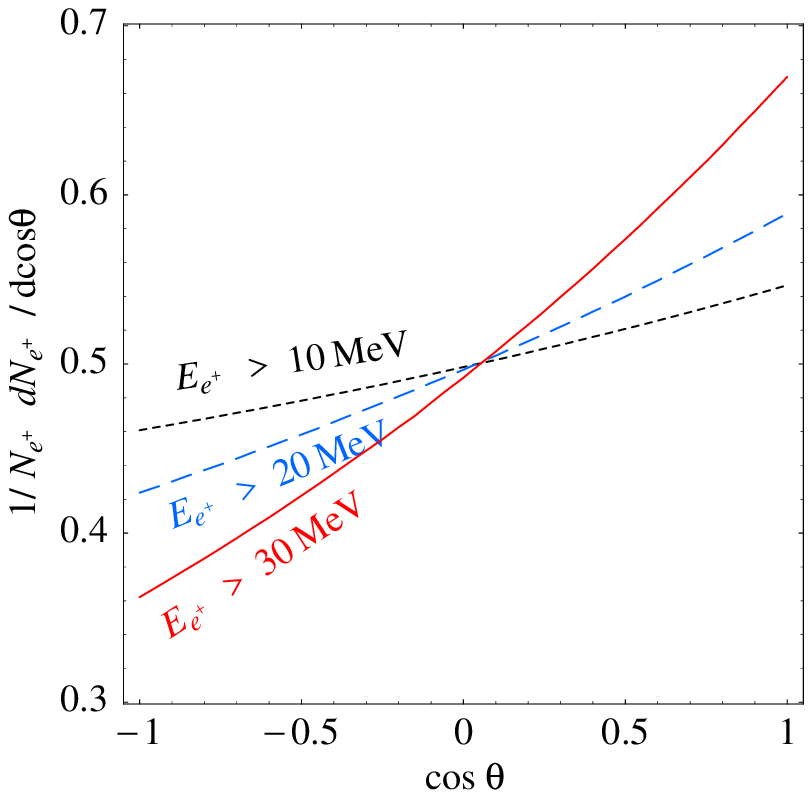}}
  \caption{\em Fig.\fig{last}a shows the energy 
spectrum of positrons. Fig.\fig{last}b shows the positron scattering 
angle distribution, for different
lower cuts on the positron energy.
We assume that supernova $\bar\nu_e$ have the
flux in eq.~(\ref{fluxsn}) and are detected via the 
$\bar\nu_e p\to n e^+$ reaction.
\label{fig:last}}
  \end{figure*}

\subsection{Distributions of positrons from supernova neutrinos}
Now we compute the energy and scattering-angle distribution
of positrons generated by supernova $\bar{\nu}_e$
and detected via the IBD cross section,
that we compute according to the formul\ae{} of
sections \ref{22} and \ref{23}. Again, we assume 
the $\bar\nu_e$ flux of eq.~(\ref{fluxsn}), where we 
set $T_e=4.5$ MeV. 
The result is shown in fig.~\ref{fig:last}.
The average cosine of the scattering angle
asymmetry is approximatively given by
\begin{equation}
\langle \cos\theta \rangle= 0.024 + 0.013\ x
\label{assehehe}
\end{equation}
($x$ as in eq.~(\ref{acc}))
which means that the scattering is slightly forward.
The forward/backward asymmetry is not particularly 
large, but could be observed in a 
large sample of neutrinos from a galactic 
core-collapse supernova using wisely 
designed selection criteria (such 
as an energy dependent angular analysis).
For example, it is sufficient to increase the 
lower cut on the positron energy to increase 
the asymmetry considerably 
\begin{equation}
\langle \cos\theta \rangle=\left\{\begin{array}{ll}
   0.028   & \hbox{for $E_e >10\,\hbox{MeV}$}   \\
   0.055   & \hbox{for $E_e >20\,\hbox{MeV}$}   \\
   0.102   & \hbox{for $E_e >30\,\hbox{MeV}$}   
\end{array}\right.
 \qquad
 \mbox{($T_e=4.5$ MeV)}.
\end{equation}
as more precisely illustrated in fig.~\ref{fig:last}b.
A lower cut on $E_e$ increases $\langle \cos\theta\rangle$ for two different reasons:
because the IBD reaction is forward peaked at higher energy, and
because forward  positrons are more energetic.
(In the Super-Kamiokande detector a lower cut $E_e \circa{>} 8\MeV$ must anyhow be imposed
in order to cut low energy backgrounds).
 Two remarks are in order:
1)  An accurate description of the direction of IBD events 
 is also important to separate other classes of events,
 as those due to reactions with Oxygen nuclei 
 in  water \v{C}erenkov detectors.
% As stressed in~\cite{vb}, the na\"{\i}ve $\bar\nu_e p$ cross section would give
% an energy-independent $\langle \cos\theta \rangle = -0.033$.
2) An accurate description of the spectrum is 
  particularly important when trying to extract information
  from the high energy tail of the distribution, about e.g.\
  oscillations in the Earth, or non-thermal corrections to the energy spectrum.

Of course, not only $\langle \cos\theta \rangle$ but also the number of events 
$N_{e^+}$ depends on the temperature.
Assuming unit efficiency above 
$E_e=8$ MeV and using eq.~(\ref{numev}), we find that 
varying $T_e$ around the reference temperature, 
the number of events varies approximately as 
$1+0.24 x -0.02 x^2$:
$N_{e^+}$ increases (decreases) by $23\%$ ($26\%$) for $T_e= (4.5\pm 1)\MeV$.

\section*{Acknowledgments}
We thank G.~Battistoni, 
V.~Berezinsky, S.~Capitani, A.~Ianni,
G.~Isidori,
H.T.~Janka, A.~Kurylov, M.Malek, D. Nicol\`o, E. Oset, O.~Palamara and K. Sato for clarifying discussions
and J.F.~Beacom and W.~Fulgione for the initial encouragement.

\footnotesize 
\frenchspacing
\begin{multicols}{2}

\end{multicols}
\end{document}